\documentclass[conference]{IEEEtran}
\IEEEoverridecommandlockouts
\usepackage{cite}
\usepackage{amsmath,amssymb,amsfonts}
\usepackage{algorithmic}
\usepackage{graphicx}
\usepackage{textcomp}

\usepackage{xcolor}
\usepackage{hyperref}
\usepackage{subfig}
\def\BibTeX{{\rm B\kern-.05em{\sc i\kern-.025em b}\kern-.08em
    T\kern-.1667em\lower.7ex\hbox{E}\kern-.125emX}}
\begin{document}

\title{3D Gaussian Splatting with Normal Information for Mesh Extraction and Improved Rendering
\thanks{MK and RD are members of PIRL (Perceptual Interfaces and Reality Lab). They were supported by ONR Award N00014-23-1-2086 and a Dolby gift.}}
\author{\IEEEauthorblockN{Meenakshi Krishnan}
\IEEEauthorblockA{Mathematics, Univ. of Maryland, \\
College Park, USA \\
mkrishn9@umd.edu} 
\and 
\IEEEauthorblockN{Liam Fowl}
\IEEEauthorblockA{Google, New York, USA\\
lfowl@google.com}
\and 
\IEEEauthorblockN{Ramani Duraiswami}
\IEEEauthorblockA{{Computer Science \& UMIACS, Univ. of Maryland} \\ College Park, USA \\
ramanid@umd.edu}}

\maketitle

\begin{abstract}
Differentiable 3D Gaussian splatting has emerged as an efficient and flexible rendering technique for representing complex scenes from a collection of 2D views and enabling high-quality real-time novel-view synthesis. However, its reliance on photometric losses can lead to imprecisely reconstructed geometry and extracted meshes, especially in regions with high curvature or fine detail. We propose a novel regularization method using the gradients of a signed distance function estimated from the Gaussians, to improve the quality of rendering while also extracting a surface mesh. The regularizing normal supervision facilitates better rendering and mesh reconstruction, which is crucial for downstream applications in video generation, animation, AR-VR and gaming. We demonstrate the effectiveness of our approach on datasets such as Mip-NeRF360, Tanks and Temples, and Deep-Blending. Our method scores higher on photorealism metrics compared to other mesh extracting rendering methods without compromising mesh quality.
\end{abstract}


\section{Introduction}
Differentiable 3D Gaussian Splatting (GS) \cite{kerbl20233d} has recently supplanted NERFs \cite{mildenhall2021nerf} as the preferred tool in rendering and reconstruction due to its low latency and accurate reconstructions. GS represents scenes using Gaussian primitives whose parameters—means, covariances, colors, and opacities—can be directly optimized using a differentiable training-free pipeline well adapted for GPU implementation, enabling real-time, highly realistic rendering without the computational expense of neural network-based volumetric renderings.

A primary application of GS is in novel-view synthesis - a technique to generate novel viewpoints of a 3D scene, given a set of 2D images. Since its introduction, it has inspired a plethora of works such as 4D Gaussian splatting for dynamic scene rendering \cite{wu20244d}, 3D content creation \cite{tang2025lgm}, human avatar modeling \cite{hu2024gaussianavatar}, and in text to 3D scene generation \cite{chen2024text}. Through view synthesis, GS-generated scenes also potentially offer a valuable source of photorealistic, synthetic images for training generative AI models, including diffusion models and GANs. It can enable practitioners to create virtually unlimited training datasets, encompassing a broad spectrum of viewpoints and scene variations. Surface mesh extraction from rendered 3D scenes is also crucial for many downstream tasks in graphics. 

In generative applications, surface mesh extraction can allow further manipulation of scenes beyond creating novel views, by aiding in editing, sculpting, and animating, all while maintaining temporal consistency. The explicit 3D representation enables object manipulation, viewpoint changes, and lighting adjustments while maintaining occlusion. Geometrically accurate rendering is also relevant in AR-VR applications and in gaming for accurate 3D scene reconstruction.

However, extracting the surface of the scene optimized via GS can be a challenge. This is because the Gaussian parameters evolve solely based on optimization of photometric losses, and may fail to accurately learn the underlying scene physical geometry. To achieve high-fidelity scene reconstruction, GS employs a densification process that significantly increases the number of Gaussians, often reaching several million 3D Gaussians with different scales and rotations, and overfits to reduce photometric error. The majority of these Gaussians are highly localized in order to enable accurate rendering of fine details and textures. This leads to a density function (whose level set describes the scene surfaces) being mostly zero everywhere, making it challenging for Marching Cubes \cite{lorensen1998marching} to generate accurate level sets, even with a high-resolution grid \cite{guedon2024sugar}.

\begin{figure*}[htbp]
  \centering
  \subfloat[]{\includegraphics[width=0.21\textwidth]{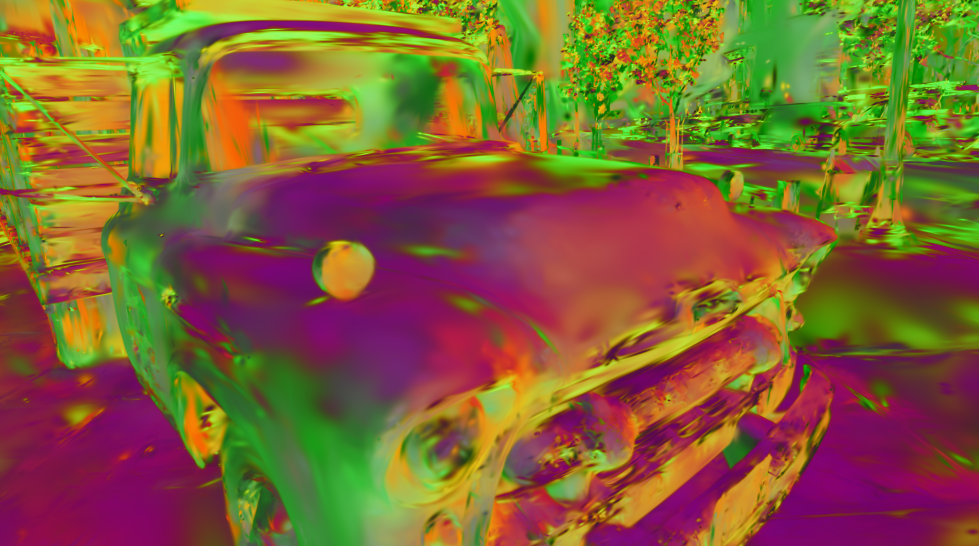}\label{fig:f1}} \hspace{1mm}
  \subfloat[]{\includegraphics[width=0.21\textwidth]{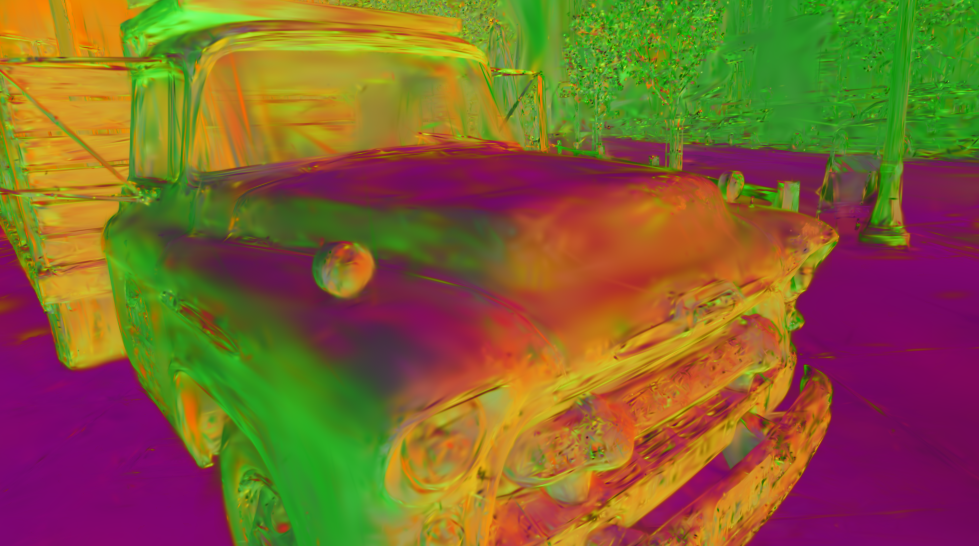}\label{fig:f2}} \hspace{1mm}
  \subfloat[]{\includegraphics[width=0.21\textwidth]{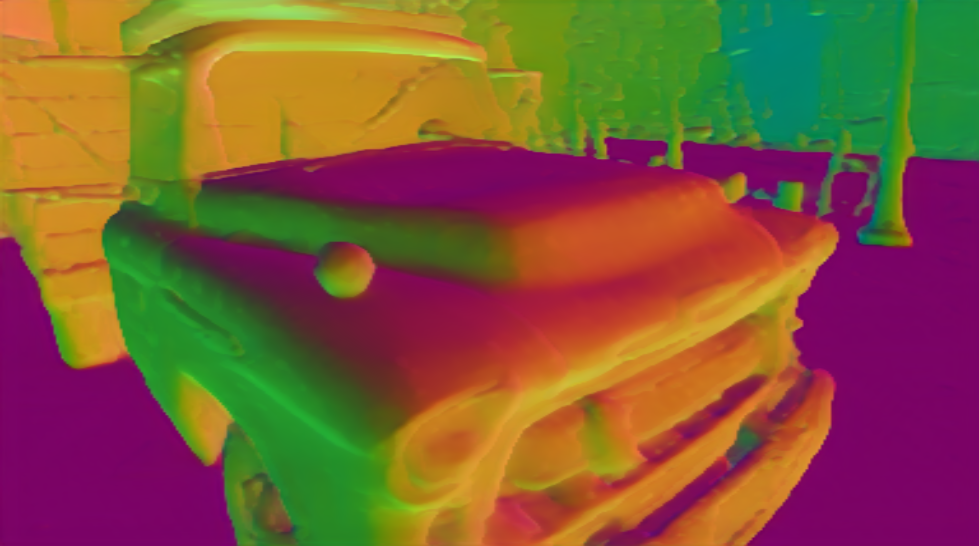}\label{fig:f3}} \hspace{1mm}
  \subfloat[]{\includegraphics[width=0.21\textwidth]{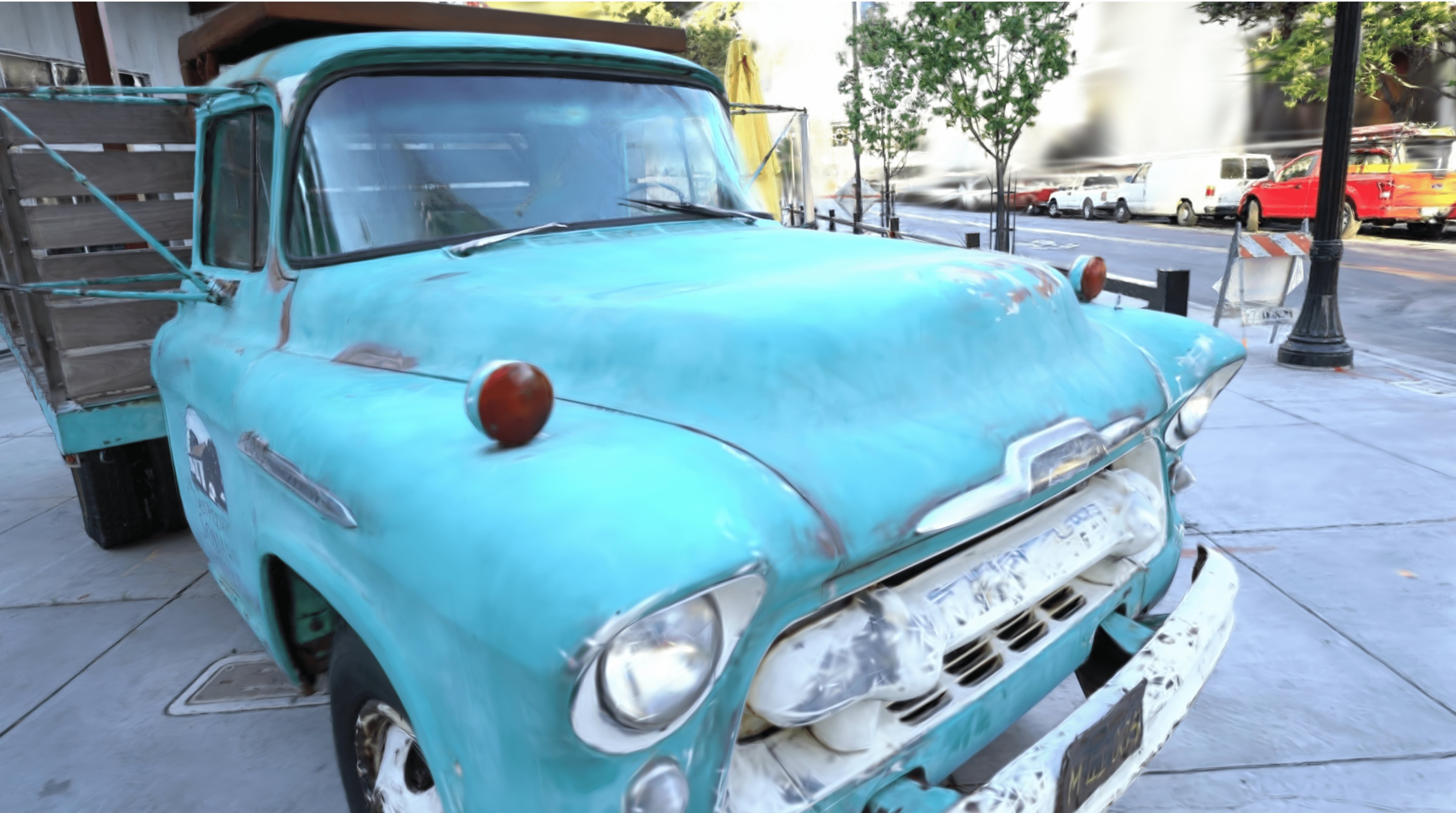}\label{fig:f4}}
  \caption{Normals before (a) and after (b) optimization with our regularizer. Also depicted are ground truth normals (c) and the final rendered image (d). Constraining the SDF gradient to be the normal enables smoother geometric transitions in GS. \label{fig:normals}}
\end{figure*}

Thus, developing effective techniques to guide Gaussian splatting for geometrically accurate rendering that enables high-quality mesh generation is of crucial importance. Our main contribution is the development of a novel regularization method which removes the reliance on heuristic target functions, and compares gradients of an estimated signed distance function (SDF) with the monocular normals generated from training images via a pre-trained neural network. Pictured in Figure \ref{fig:f1} is the effectiveness of our approach in guiding these gradients, resulting in much smoother, less noisy normals. Our method simultaneously produces scene representations that facilitate easy mesh extraction, while also improving the photorealism of novel views of the scene compared to existing works that also facilitate mesh generation for GS, and is demonstrated in several experiments.

\section{Related Work}
\subsection{Neural Rendering} Several approaches have applied deep learning techniques to the novel-vew synthesis problem to varying degrees of success \cite{flynn2016deepstereo, zhou2016view, hedman2018deep}. Perhaps the most well known deep learning approach, Neural Radiance Fields (NeRF) \cite{mildenhall2021nerf}, has garnered significant attention in the field of 3D scene reconstruction for its high quality rendering. In this method, 3D scenes are represented as continuous volumetric functions using multi-layer perceptrons (MLPs) which takes as inputs 3D positions
and viewing directions, and outputs density and view-dependent colors
for differentiable rendering. NeRF's impressive rendering results come at the cost of higher latency due to the inclusion of large MLP layers. Several follow-up works have modified NeRF's strategy to improve rendering quality, or decrease training time, with features like regularization terms \cite{niemeyer2022regnerf}, data-structures \cite{chen2022tensorf}, or different encoders \cite{barron2021mip,muller2022instant}. 

Using a neural representation to parametrize the SDF of a 3D scene has also been the subject of various studies \cite{wang2021neus, yu2022monosdf}. Normal supervision using ground-truth normals \cite{gropp2020implicit}, Eikonal loss \cite{ben2022digs} or incorporating properties of the SDF function such as the eigenvectors of its Hessian \cite{wang2024aligning} as regularizing terms has recently gained attention as well. Our goal is to extend this approach to 3D GS.

\subsection{Point-Based Rendering} 

Point-based rendering techniques represent 3D scenes as dense clouds of points, rather than polygons, relying on points to uncover underlying surface details. Such methods represents a different approach to the problem of novel-view synthesis that does away with deep learning pipelines and often instead focuses on other parametric approaches. In \cite{ruckert2022adop}
a rasterizer renders points as one-pixel splats while a sphere-based scene representation is used in \cite{Lassner_2021_CVPR}. Differentiable 3D GS \cite{kerbl20233d}, based on the ideas proposed in \cite{zwicker2002ewa}, chooses Gaussian primitives whose parametrization naturally admits a differentiable volume representation, while also allowing fast rendering via $\alpha$-blending. Various works have suggested further improvements such as anti-aliased rendering \cite{yu2024mip}, improving the capability
of rendering view-dependent effects \cite{lu2024scaffold}, and compression and regularization \cite{lee2024compact,girish2023eagles}.
Other modern non deep learning approaches have also achieved impressive results producing photorealistic novel views. One popular example in the same vein as GS is plenoxels \cite{fridovich2022plenoxels} which optimizes a spherical harmonics representation of a scene. 
\subsection{Mesh Representation}
Another way of representing a scene is with a \textit{mesh} that accurately describes surfaces and objects in it. As previously detailed, mesh generation is crucial for several applications of scene rendering. However, both GS and NeRFs can struggle with mesh generation - but for different reasons. 

For NeRFs, several mesh extraction techniques have been proposed, such as by embedding opacities and features into texture maps \cite{chen2023mobilenerf} or baking the neural fields into textures\cite{yariv2023bakedsdf}. For the purposes of this work, we will focus on mesh extraction for GS. Since the standard 3D GS method does not explicitly encode any geometric information into the structure of the learned Gaussians, several works have tried to encode more geometric information into the GS process \cite{guedon2024sugar, turkulainen2024dn, zhang2021physg}. 

A distance-based Gaussian splatting technique for better mesh extraction was introduced in \cite{choi2024meshgs}, while \cite{waczynska2024games} suggests tightly binding the 3D Gaussian splats to
meshes. In \cite{guedon2024sugar}, a complex regularization method and an alternate strategy to extract meshes from the Gaussians were proposed. However, this often introduces \textit{several} complex regularization terms - each with their own associated hyperparameters. It also comes at the cost of diminished performance on the photorealism front. 
One way to encode more geometric structure into the GS process is with \textit{normal supervision}. Gao et al \cite{gao2023relightable} propose using the gradients of rendered depths as pseudo-ground-truth for normal supervision. Closely related to our work is \cite{turkulainen2024dn} where neural network estimated monocular normals are used to supervise the smallest scaling axis of a Gaussian that is encouraged to be flat to align with the surface.

\section{Method}
We follow the general pipeline as in the original Gaussian Splatting paper \cite{kerbl20233d}. Like NeRF methods, the inputs are cameras calibrated with Structure-from-Motion (SfM) \cite{schonberger2016structure}, and Gaussians are initialized using the sparse point cloud obtained. The 3D scene is represented by a large set of Gaussians, each parametrized by its mean $\mu$, and covariance $\Sigma$:
\[
G(x,y) = \exp \left(-\frac12 (x-\mu)^T \Sigma^{-1} (x-\mu)\right).
\]
They are also associated with opacity coefficients $\alpha \in [0,1]$ and spherical harmonics coefficients that represent the colors emitted by the Gaussians in all directions. These parameters are optimized during the training process. To ensure that the positive semi-definite property of the covariance matrices is retained during the optimization process, these matrices are parametrized using rotation and scaling matrices denoted as $R$ and $S$, respectively with $\Sigma = RSS^TR^T$.

The 3D Gaussians are projected onto the 2D image plane as splats and blended using weighted sums of color and opacity. A fast tile-based rasterization method efficiently does the 3D-to-2D projection and applies $\alpha-$blending which gives GS its main speed advantage over neural volumetric representations. The optimization is interleaved with adaptive density control of the Gaussians where more are added to both ``empty" areas (with missing geometric features) and areas where a single Gaussian covers large parts of the scene. Essentially transparent Gaussians with opacities close to 0 are pruned between iterations.

For a given GS scene, \cite{guedon2024sugar} defines a corresponding density function at any space location $p$ as sum of Gaussian values weighted by their alpha-blending coefficients,
\begin{align}\label{eqn:density}
    d(p) = \sum_g \alpha_g  \exp \left(-\frac12 (p-\mu_g)^T \Sigma_g^{-1} (p-\mu_g)\right).
\end{align}
They then propose a target density function that promotes flat and surface-aligned optimized Gaussians to supervise (\ref{eqn:density}). To further increase surface alignment, they also define an ideal SDF function whose zero level sets correspond to the surface of the scene. To estimate the SDF for the current iteration, the depth maps of the Gaussians
from the training viewpoints are used which can be rendered using the splatting rasterizer. Then for a point $p$ visible from a training viewpoint, the estimated SDF $\hat{f}(p)$ at that point is taken to be the difference between the depth of $p$ and the depth in the corresponding depth map at the
projection of $p$. Additional regularization is applied for opacity, and the SDF gradient is pushed towards the direction of the smallest axis of the Gaussian.

Instead of a target SDF, we propose supervising the gradients of the SDF using its (unsigned) cosine similarity with the monocular normals obtained from a pretrained neural network such as Omnidata \cite{eftekhar2021omnidata} or MiDaS \cite{ranftl2020towards}. First the gradients are projected into the camera space using the world-to-view transform, after which their values at the corresponding pixel positions are obtained using the GS rasterizer. Specifically, at a point $p$, denoting the rasterized gradient as $\nabla \hat{f}_p$ and the normals as $n_p$, we have a regularizing term of the form:
\begin{align}
    R(\hat{f}) = \sum_{p \in P} \frac{|\nabla \hat{f_p} \cdot n_{p}|}{\|\nabla \hat{f}_p\| \ \|n_p\| }.
\end{align}
Here $P$ denotes the set of pixel points in the training images. Then for the regularizing iterations, the loss is taken to be
\begin{align*}
    \mathcal{L} = \mathcal{L}_1 + \lambda_{\rm DSSIM}  \mathcal{L}_{\rm DSSIM} + \lambda_{r}  R(\hat{f}),
\end{align*}
where $\mathcal{L}_1$ and $\mathcal{L}_{\rm DSSIM}$ denote the standard $L^1$ norm loss and the D-SSIM term. We set $\lambda_{\rm DSSIM} = 0.2$ and $\lambda_{r}$ is tuned for the different scenes; a value between 0.1 and 0.4 works well. Additionally, we use entropy regularization for a short number of iterations to encourage the opacities to be binary, since we need the Gaussians to represent a surface and want to avoid semi-transparent ones. A mesh is extracted from the Gaussians obtained after optimization with the regularization terms by running a Poisson reconstruction \cite{kazhdan2006poisson} on points sampled on a level set of the trained SDF function, as in \cite{guedon2024sugar}. Further refinement of the Gaussians can also be employed to improve the mesh quality by initializing a new set of Gaussians on the centers of the coarse mesh and jointly tuning the Gaussian as well as the mesh representation, which can end up being more computationally expensive. For our approach, we demonstrate that even with a shorter refining time, the mesh quality and rendering quality are not compromised when we run our computationally efficient regularizing for more iterations.

\section{Results and Discussion}

\subsection{Implementation Details} All experiments were done using a NVIDIA RTX A6000 GPU. We follow the initial training regime in \cite{guedon2024sugar}, performing 7000 iterations of vanilla GS, followed by 2000 iterations of opacity regularization. After this, we do either 6000, or 13000 iterations with our regularization. Depending on the number of regularizing iterations, we then have a refinement stage of either 15000 or 7000 iterations for a total of 30000 iterations. We use Omnidata \cite{eftekhar2021omnidata} to generate the monocular normals.

\subsection{Results} We compare the PSNR, SSIM, LPIPS \cite{zhang2018unreasonable} scores for various rendering datasets. Experiments were performed on the Tanks\&Temples dataset \cite{knapitsch2017tanks} on the scenes \textit{truck} and \textit{train}. We also present results on the synthetic \textit{Dr Johnson} and $\textit{Playroom}$ datasets from the Deep Blending repository \cite{hedman2018deep}. Finally, we evaluate performance on realistic scenes from the MipNeRF360 \cite{barron2021mip}, thus covering both indoor and outdoor scenes. We compare both with methods whose goal is to optimize rendering, and methods that extract meshes.

\begin{table}[]\centering \caption{Results on the MipNerf Dataset for indoor and outdoor scenes with baselines reported in \cite{guedon2024sugar}.\label{tab:mipnerf}
}
\renewcommand{\arraystretch}{1.2}
\begin{tabular}{|c|ccc|}
\hline
\textbf{Indoor} & PSNR $\uparrow$ & SSIM $\uparrow$ & LPIPS $\downarrow$ \\ \hline
Plenoxels \cite{zhang2021physg} & 24.83 & 0.766 & 0.426 \\
INGP-Base \cite{muller2022instant} & 28.65 & 0.840 & 0.281 \\
INGP-Big \cite{muller2022instant} & 29.14 & 0.863 & 0.242 \\
Mip-NeRF360 \cite{barron2021mip} & 31.58 & 0.914 & 0.182 \\
Baked-SDF \cite{yariv2023bakedsdf} & 27.06 & 0.836 & 0.258 \\
3DGS \cite{kerbl20233d} & 30.41 & 0.920 & 0.189 \\
SuGaR \cite{guedon2024sugar} & 29.43 & 0.910 & 0.216 \\
Ours & 29.87 & 0.921 & 0.117\\
\hline
\end{tabular}
\\
\begin{tabular}{|c|ccc|}
\hline
\textbf{Outdoor} & PSNR $\uparrow$ & SSIM $\uparrow$ & LPIPS $\downarrow$ \\ \hline
Plenoxels \cite{zhang2021physg} & 22.02 & 0.542 & 0.465 \\
INGP-Base \cite{muller2022instant} & 23.47 & 0.571 & 0.416 \\
INGP-Big \cite{muller2022instant} & 23.57 & 0.602 & 0.375 \\
Mip-NeRF360 \cite{barron2021mip} & 25.79 & 0.746 & 0.247 \\
Mobile-NeRF \cite{chen2023mobilenerf} & 21.95 & 0.470 & 0.470 \\
3DGS \cite{kerbl20233d} & 26.40 & 0.805 & 0.173 \\
SuGaR \cite{guedon2024sugar} & 24.40 & 0.699 & 0.301 \\
Ours & 24.61 & 0.719 & 0.262\\
\hline
\end{tabular}
\end{table}

\begin{table}[]\centering \caption{Results on the Tanks\&Temples dataset with reported baseline values\cite{guedon2024sugar}.\label{tab:tanks}}
\renewcommand{\arraystretch}{1.2}
\begin{tabular}{|c|ccc|}
\hline
 & PSNR $\uparrow$ & SSIM $\uparrow$ & LPIPS $\downarrow$ \\ \hline
Plenoxels \cite{zhang2021physg} & 21.07 & 0.719 & 0.379 \\
INGP-Base \cite{muller2022instant} & 21.72 & 0.723 & 0.330 \\
INGP-Big \cite{muller2022instant} & 21.92 & 0.744 & 0.304 \\
Mip-NeRF360 \cite{barron2021mip} & 22.22 & 0.758 & 0.257 \\
3DGS \cite{kerbl20233d} & 23.14 & 0.841 & 0.183 \\
SuGaR \cite{guedon2024sugar} & 21.58 & 0.795 & 0.219 \\
Ours & 21.83 & 0.802 & 0.216\\
\hline
\end{tabular}
\end{table}

\begin{table}[]\centering \caption{Results on the Deep Blending dataset with reported baseline values \cite{guedon2024sugar}.\label{tab:deepblend}}
\renewcommand{\arraystretch}{1.2}
\begin{tabular}{|c|ccc|}
\hline
 & PSNR $\uparrow$ & SSIM $\uparrow$ & LPIPS $\downarrow$ \\ \hline
Plenoxels \cite{zhang2021physg} & 23.06 & 0.794 & 0.510 \\
INGP-Base\cite{muller2022instant} & 23.62 & 0.796 & 0.423 \\
INGP-Big\cite{muller2022instant} & 24.96 & 0.817 & 0.390 \\
Mip-NeRF360 \cite{barron2021mip} & 29.40 & 0.901 & 0.244 \\
3DGS \cite{kerbl20233d}& 29.41 & 0.903 & 0.242 \\
SuGaR \cite{guedon2024sugar} & 29.41 & 0.893 & 0.267 \\
Ours & 29.55 & 0.893 & 0.269 \\
Ours (7K refinement) & 29.72 & 0.894 & 0.277 \\
\hline
\end{tabular}
\end{table}

Note that in our experiments, GS sometimes performs better than our method in terms of photorealism metrics. This is not unexpected, and we emphasize that our goal is not necessarily to beat vanilla GS in terms of these metrics, but instead to produce scene representations that simultaneously perform well photorealistically while \textit{also} being good for mesh generation - a feature lacking in standard GS. The closest work for comparison is SuGaR \cite{guedon2024sugar} which also proposes a modification to vanilla GS for better mesh generation. 

\begin{figure}
  \centering
  \subfloat[]{\includegraphics[width=0.23\textwidth]{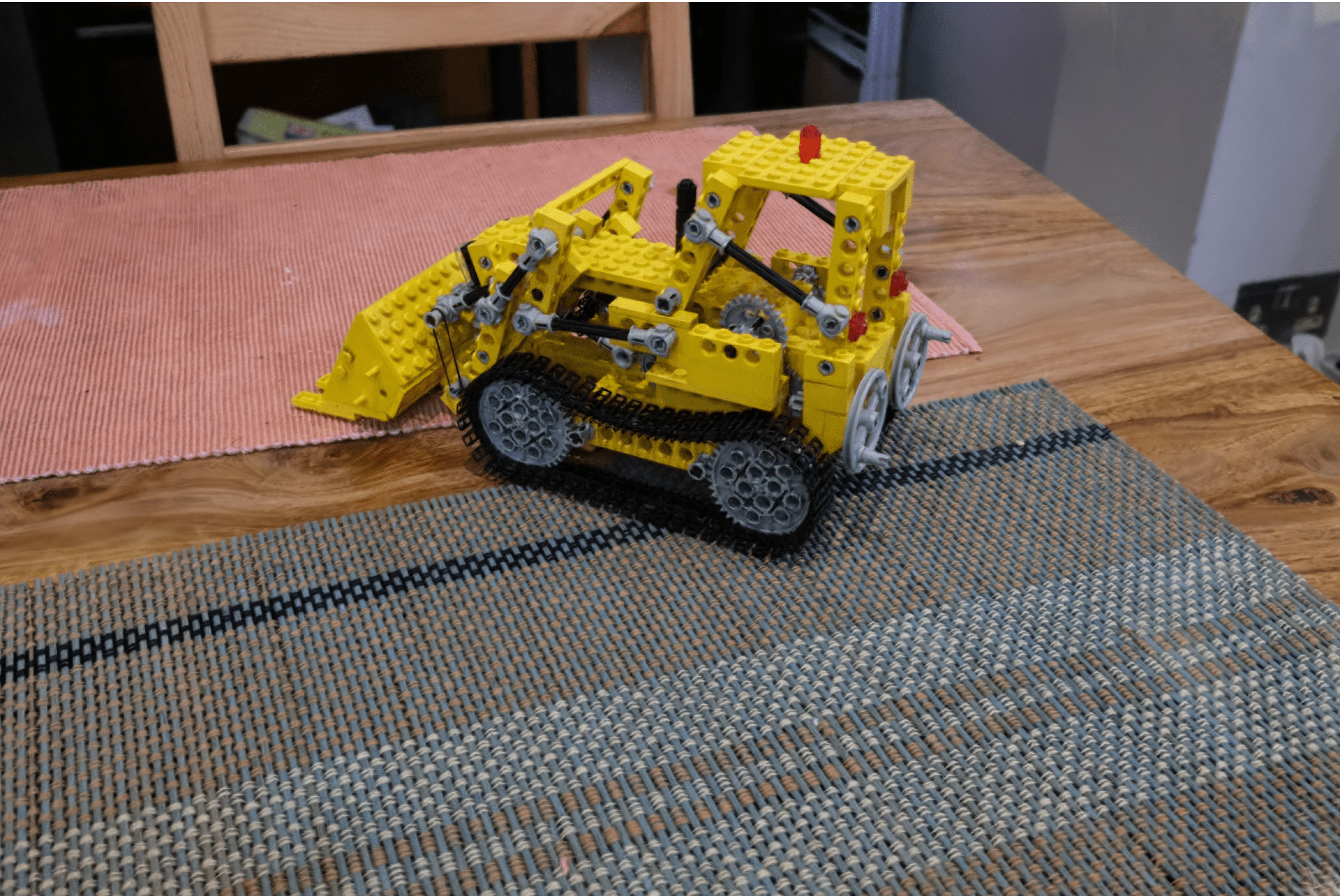}\label{fig:f5}}
  \hspace{1mm}
  \subfloat[]
  {\includegraphics[width=0.23\textwidth]{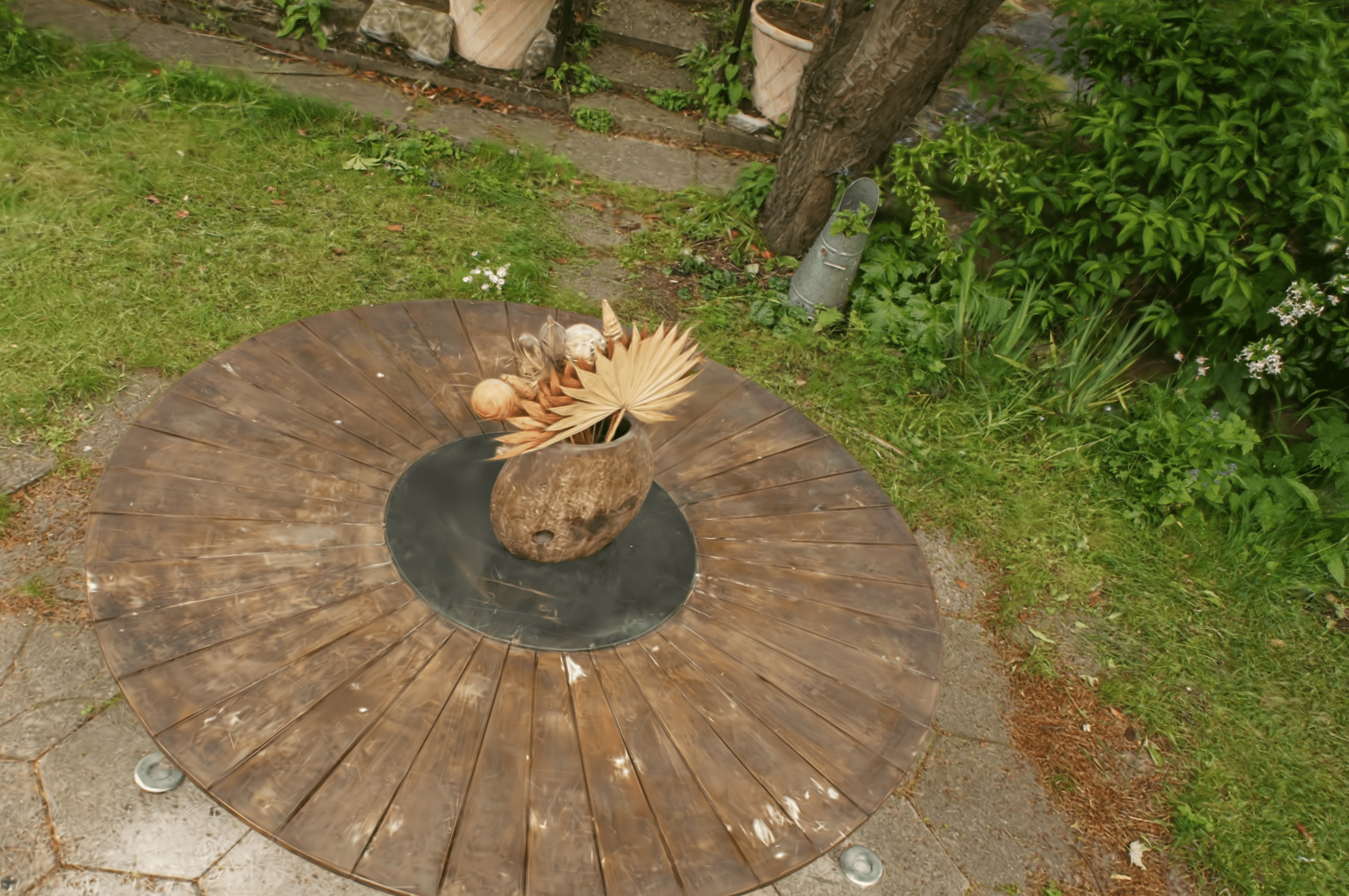}\label{fig:f6}}
  \caption{Rendered images using our regularization. \label{fig:render}}
\end{figure}

Figure \ref{fig:render} demonstrates some of the images rendered using our method. For the MipNerF dataset we demonstrate competitive results with SoTA methods in rendering quality, and improve over the mesh-extracting methods (SuGaR, Baked-SDF, Mobile-NeRF) for both indoor and outdoor scenes. We report results for seven scenes from this dataset for our tests (all besides \textit{Flowers} and \textit{Treehill}) in Table \ref{tab:mipnerf}. For the Tanks\&Temples dataset, our regularization improves over SuGaR in all metrics (Table \ref{tab:tanks}). Our regularization performs better than \textit{all} other baselines for the Deep Blending dataset. On this dataset, we achieve a higher average PSNR than state-of-the-art methods (see Table \ref{tab:deepblend}). We see that increasing the number of regularization iterations improves the quality of rendering but at the cost of a lower quality mesh for this dataset. 

\begin{figure}
  \centering
  \subfloat[]{\includegraphics[width=0.245\textwidth]{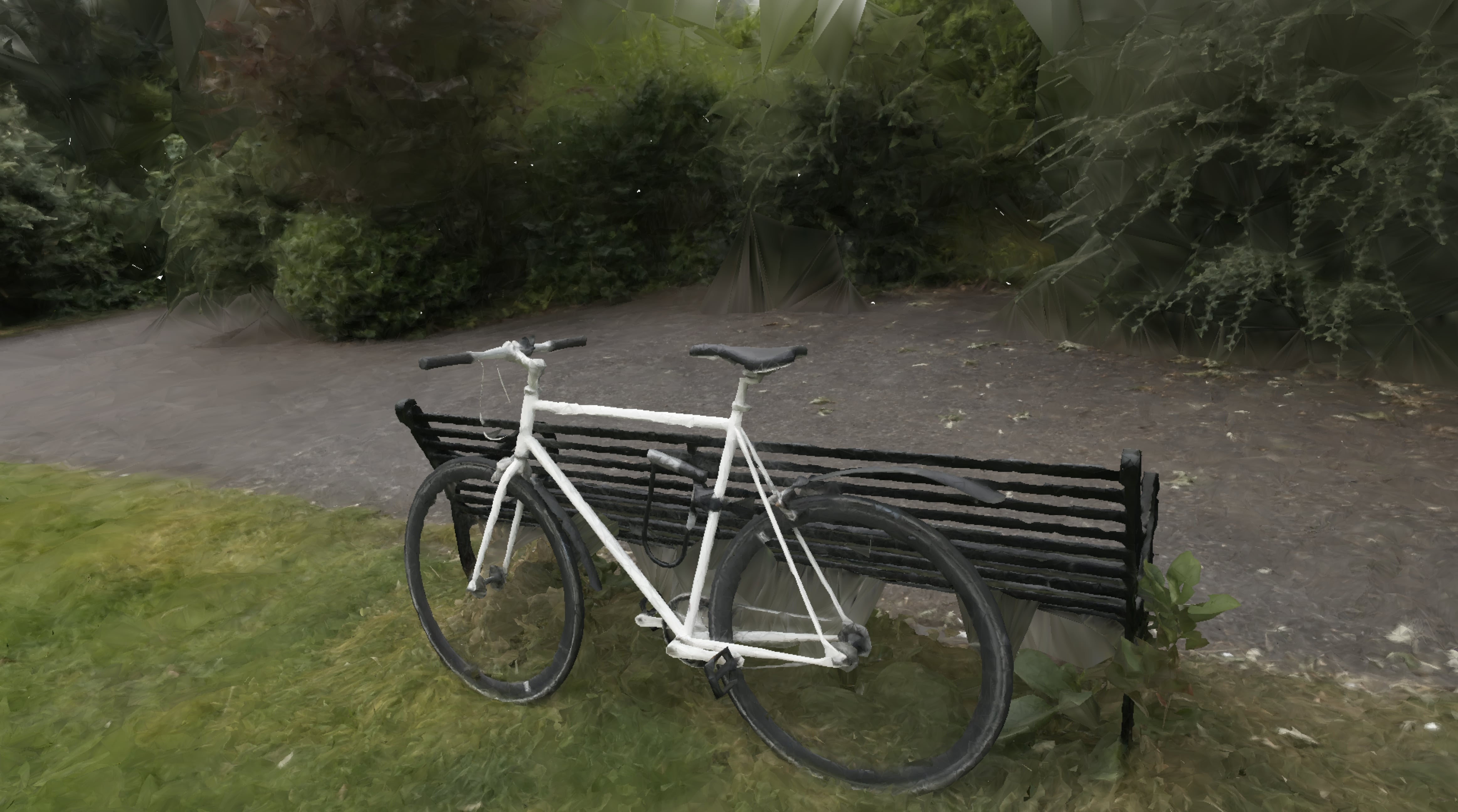}\label{fig:f7}}\hspace{1mm}
  \subfloat[]
  {\includegraphics[width=0.205\textwidth]{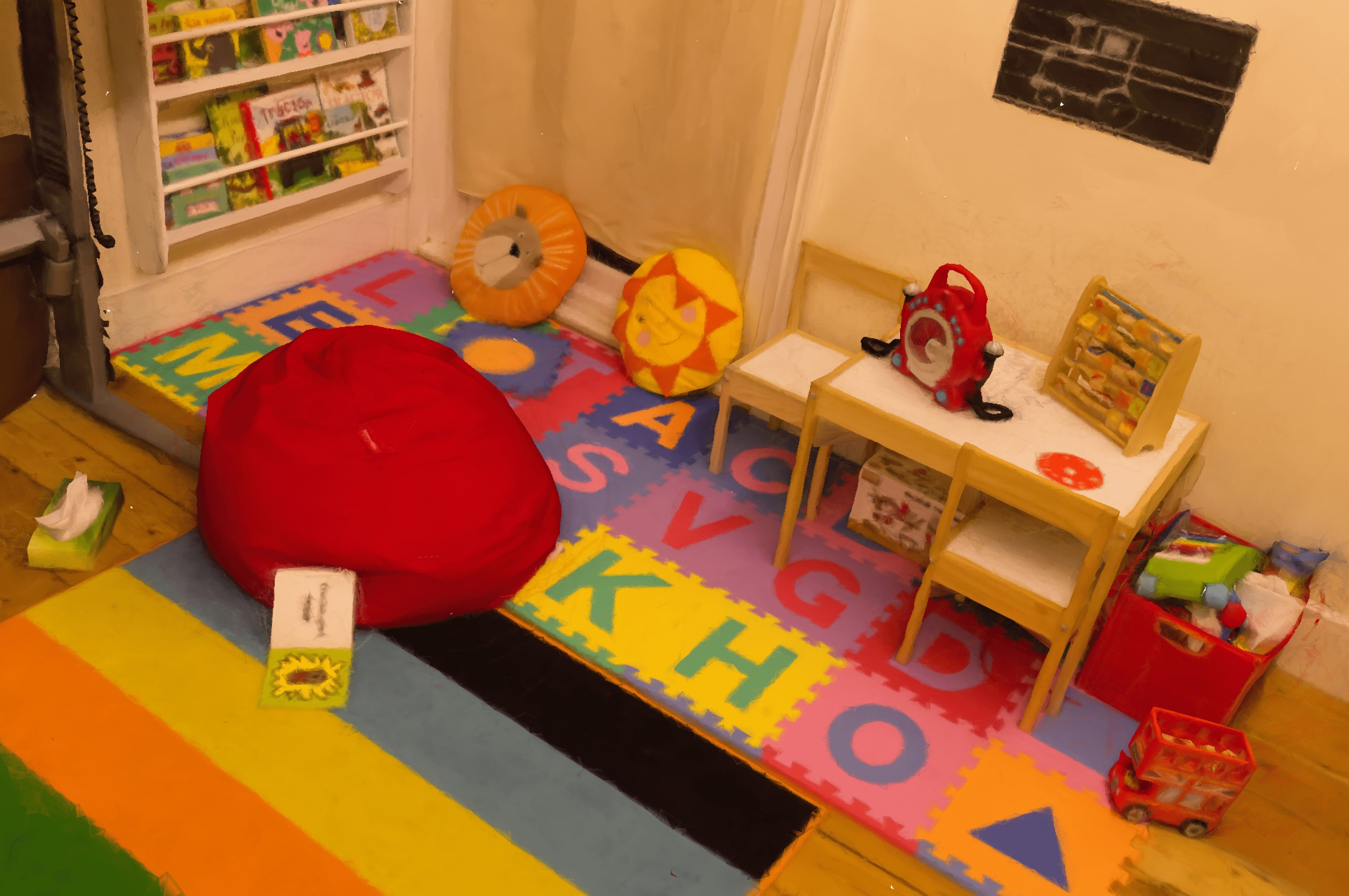}\label{fig:f9}} \hspace{1mm}
  \caption{UV-Textured Meshes extracted after optimizing with our regularizing term, with colors rasterized onto the mesh. \label{fig:mesh}}
\end{figure}

\begin{figure}
  \centering
    \subfloat[]{\includegraphics[width=0.23\textwidth]{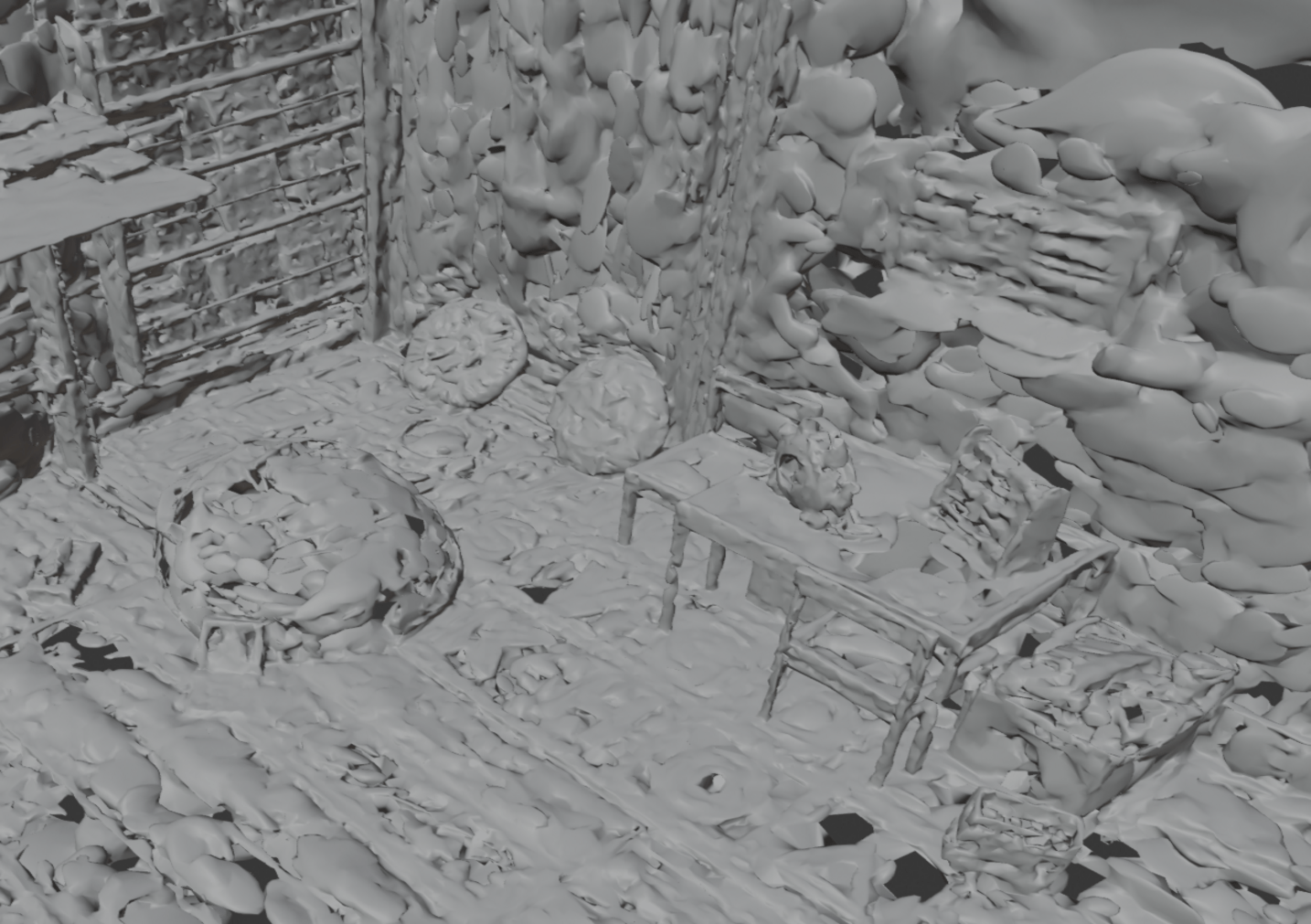}\label{fig:f8}}\hspace{1mm} 
    \subfloat[]{\includegraphics[width=0.23\textwidth]{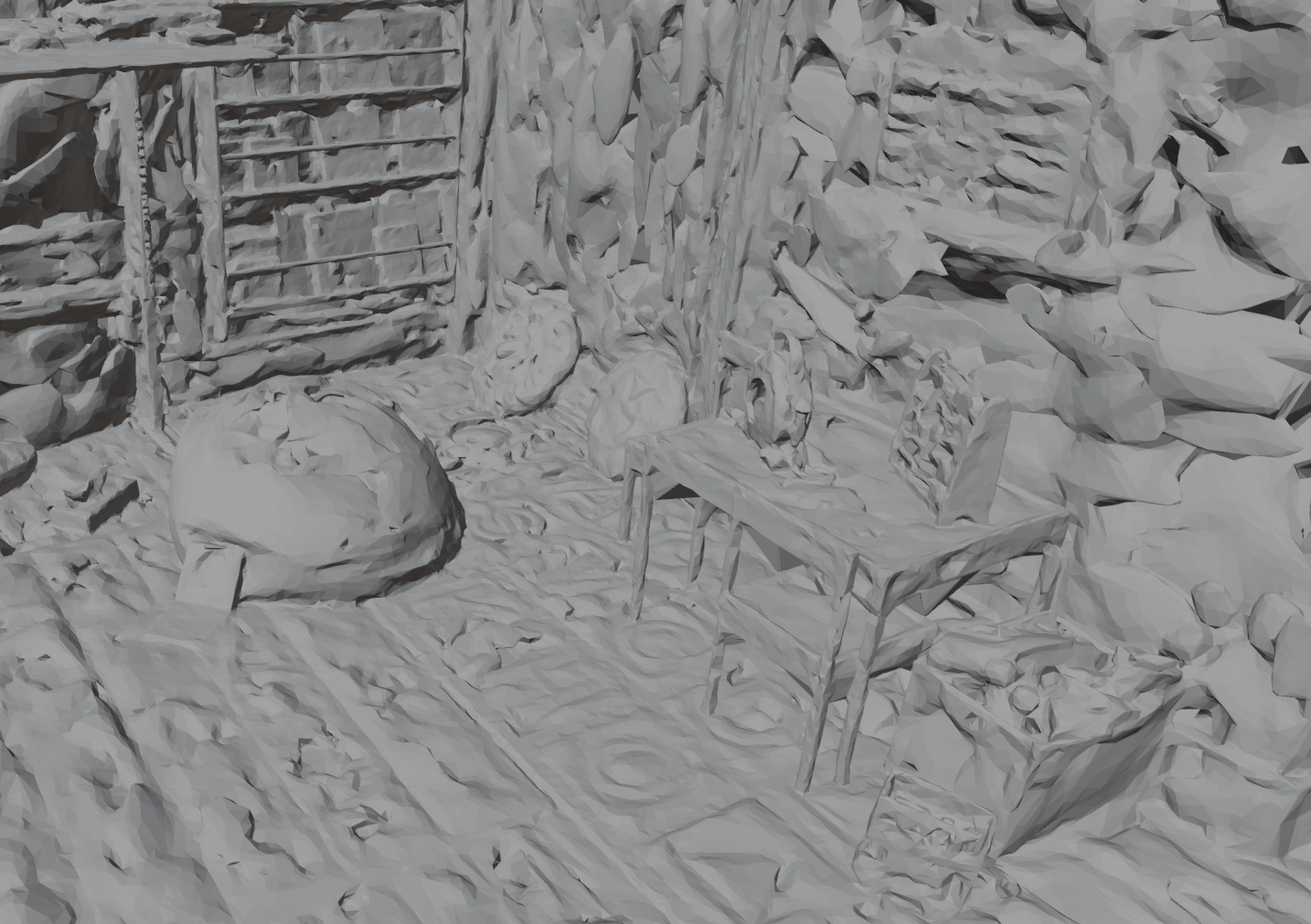}\label{fig:f8}}
    \caption{Textureless meshes extracted from Gaussians (a) for vanilla GS and (b) after optimizing with our regularizing term for the scene in \ref{fig:f9}. The 3DGS mesh contains holes, and is very noisy. \label{fig:mesh1}}
\end{figure}

Figures \ref{fig:mesh} and \ref{fig:mesh1} illustrate the UV-textured meshes extracted from the scenes, with triangle faces colored using the GS rasterizer, and the untextured meshes extracted from our method and vanilla GS, respectively. Both methods employ the same mesh-extraction technique from \cite{guedon2024sugar}, but the vanilla GS mesh is noticeably noisy and contains holes. In Figure \ref{fig:normals}, we show the alignment of the estimated normals before and after adding the regularizing term. Clearly the method is able to denoise the normal estimation, and guide the Gaussians towards smoother geometric transitions.

That the rendering quality of our method can surpass that of 3D GS for some scenes is unsurprising because the regularizing term helps to place and align the Gaussians along regions of fine detail and texture.



\begin{figure}[htbp]
  \centering
  \subfloat[]{\includegraphics[width=0.21\textwidth]{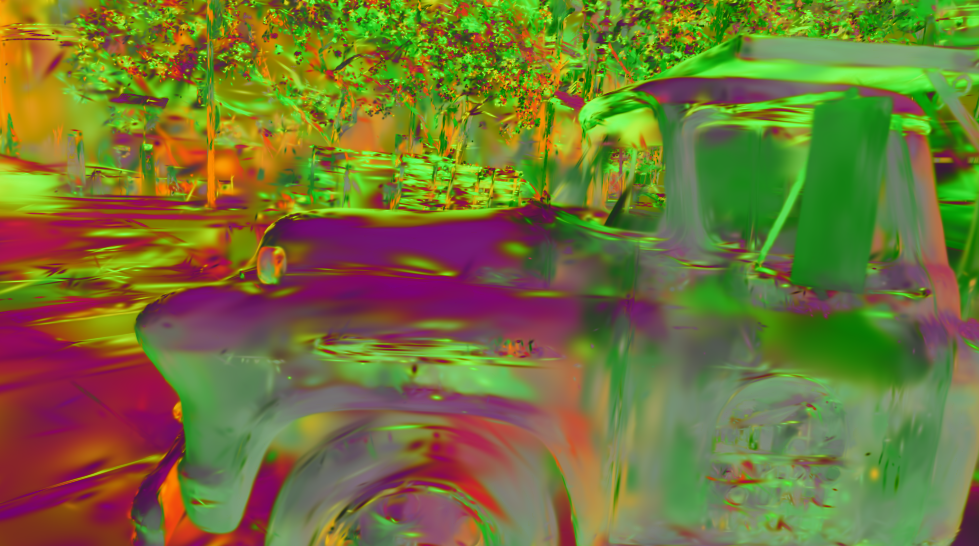}\label{fig:f10}} \hspace{1mm}
  \subfloat[]
  {\includegraphics[width=0.21\textwidth]{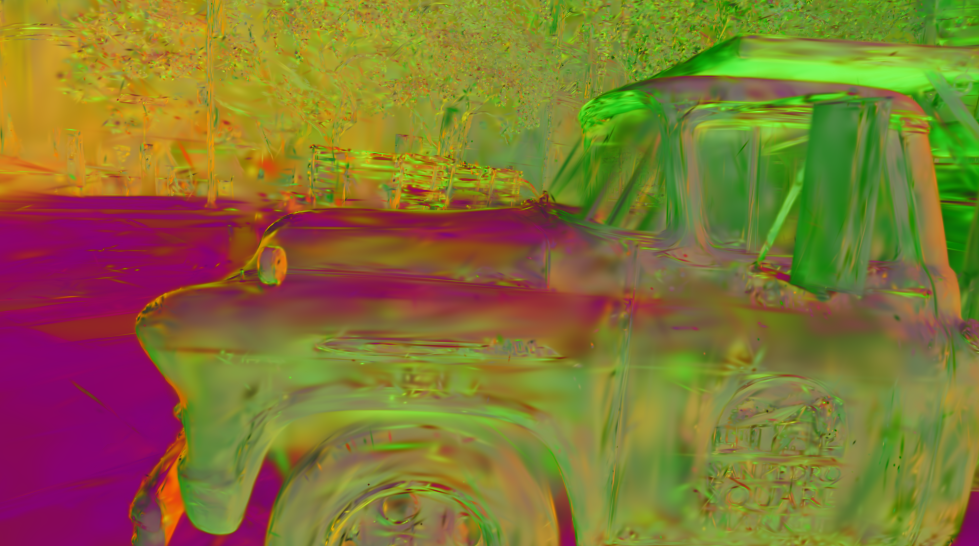}\label{fig:f11}}
  \caption{Computed normals without and with regularization. \label{fig:normals}}
\end{figure}

\section{Conclusion}
We introduce a novel regularization term for Gaussian splatting for geometrically accurate rendering and surface mesh extraction. We use a pre-trained neural network to predict per pixel monocular normals, that are then used to supervise the gradient of an estimated signed distance function whose zero level sets are the scene surface. We demonstrate improvement in the rendering quality while also extracting a surface mesh of the 3D scene.

\newpage

\bibliographystyle{IEEEtran}
\bibliography{refs}


\end{document}